\documentclass[mmnp]{edpsmath}
\usepackage{graphicx,epsfig,color}
\begin{document}
\selectlanguage{english}
\title{On the shape of invading population in anisotropic environments}

\author{Viktoria Blavatska}\address{Department for Computer Simulations of Many-Particle Systems, \\
              Institute for Condensed Matter Physics
               of the National Academy of Sciences of Ukraine,\\
              Lviv 79011, Ukraine}

\date{\today}
\begin{abstract} We  analyze the properties of population spreading in environments with spatial anisotropy 
within the frames of a lattice model of asymmetric (biased) random walkers.  
The expressions for the universal shape characteristics of the instantaneous configuration of 
population, such as asphericity $A$ 
and prolateness $S$ are found analytically and proved to be dependent only 
on the asymmetric transition probabilities in different directions. The model under consideration 
is shown to capture, in particular, the peculiarities of invasion in presence of an array of oriented tubes (fibers) in the environment. \end{abstract}
%
%
\maketitle
\section*{Introduction}
The problem of spreading of a population of agents in anisotropic environments with some preferred 
orientations of movement  is 
encountered in a rich variety of biological phenomena and medicine. In general, the movement characteristics  are
influenced  due to the environmental heterogeneity.
Important examples  include 
 behavior of chemosensitive cells like bacteria or leukocytes in the
gradient of a chemotactic factor \cite{Alt80, Berg00, Chalub04, Dolak05}, the motion of micro-organisms 
under the action of gravitational force (gravitaxis) or a light source (phototaxis) \cite{Hill97, Pedley90, Vincent96}, 
the cell migration in fiber network of extracellular environment
 \cite{Chauviere07, Dickinson94, Dickinson00, Hillen06,  Painter09} and in particular 
the tumor invasion and metastasis in the tissue matrix \cite{Alarcon03, Anderson00, Friedl03}. 
The important problems of spatial ecology are connected with analysis of migrations of living
 organisms in oriented habitats with orientation 
 given by magnetic cues, elevation profiles, spatial distributions of resources \cite{Armsworth05, Dewhirst09, Hillen13}.
 The variations of above  characteristics has non-trivial influence on the population growth, persistence and 
 dispersal \cite{Armsworth05, Cantrell06, Kirk97, Yurk17}. 
  In particular, the presence of oriented factors in environment lead to occurrence of directed movement patterns, 
 different from pure diffusion \cite{Armsworth99, Belgacem95, Cantrell06}.
 In this concern,
it is worthwhile to mention also the modern technologies of controlled drug delivery, 
using an external oriented magnetic field  \cite{Zakharchenko18},
and  the implants
based on arrays of oriented TiO$_2$ nanotubes, which control  
the directed release of drugs \cite{Losic15, Wang17}.  

The model of a random walk (RW) on a regular  lattice  provides a good description of
the stochastic  processes \cite{RWbook}. In the simplest case, 
when there is no preferred direction, this process restores the Brownian
motion and such a model may be shown to produce the
standard diffusion equation. 
Making the probabilities of moving in different directions not equal causes the directional bias, 
which leads to the drift-diffusion  equation. Such asymmetric biased random walks (BRW) 
are frequently used in biology to model the motion of living organisms and 
cells in oriented environments \cite{Codling17, Patlack53}. 
The bias may be caused both by the fixed external environmental factors (such as gravitational force or external magnetic field), 
and by varying factors (such as chemical gradient or food resources in oriented migration of organisms). 
Thus, the transition probabilities   
 in BRW model can also be not only constants, but also functions of space and time. 

{

The problem of determining the size and shape of individual random walk trajectory, treated as track of a particle (cell) in environment, 
attracts a considerable attention of researchers. 
In the pioneering paper of Kuhn \cite{Kuhn34} it was suggested, 
that RW trajectory is a highly aspherical object. In the following studies  \cite{Solc71, Solc71a}  the averaged principal components of inertia tensor were introduced       
as parameters for description of RW asymmetry.  Later in Refs. \cite{Aronovitz86, Rudnick86} it was proposed to characterize the shape properties by the set of rotationally invariant combinations of inertia tensor components, such as asphericity $A$ and prolateness $S$. These shape characteristics of RW trajectories
were estimated both analytically 
 \cite{Diehl89, Wei97, Gaspari87, Sciutto94} and numerically  \cite{Bishop88}.  

From the point of view of biological application,
an importance of analyzing the tracks of single
cells can be realized  e.g. in the processes of the guidance of dendritic cells by haptotactic
chemokine gradients towards lymphatic vessels \cite{Weber13} or for neutrophil migration directed by inflammatory
chemokines \cite{Sarris12}. In particular,  the neutrophil migration
appeared to be of random walk type with directed track segments induced 
by chemokine gradients. 
The universal shape parameters, such as asphericity of individual cell 
tracks were evaluated in  \cite{Mokhtari13} making use of image data  
for synthetic cells and {\it in vitro} neutrophil tracks, obtained in microscopy experiments.  

{ Processes of  bacteria growing and invasion are known to produce colonies of various shapes, called ``patterns" or ``morphotypes" \cite{Shapiro95,Rudner98,Franco02}. 
	Colony patterns serve for differentiation of populations of individuals otherwise identical. 
	The shape of a bacterial colony  depends on variables like the nutrient diffusion field, 
	 cyclic production of chemoattractants and chemorepellents, long range chemical signalling such as quorum sensing and production of secreted wetting fluid \cite{Jacob04}.
	The significance of analyzing the bacterial colony patterns
	resides in a deeper understanding of colony evolution and morphogenesis in the given environment.
	
	Different technologies like  genetic engineering and histochemical staining \cite{Shapiro84},  scanning electron microscopy \cite {Shapiro85} etc. are applied 
	to investigate the  geometry of growth and invasion of bacteria colonies. 
	The asymmetric shape pattern formation is found e.g. when colonies encountered obstacles in substrate during development, such as glass fibers or other bacteria co-existing \cite{Shapiro95,Rudner98}
	or chemical fields \cite{Salhi93}. Formation of patterns of unusual shape  can also signalize about arising of mutations in given population \cite{Franco02,Mendelson76}.
	Processes of  cell migration in disordered extracellular environment and  tissue matrices
	\cite{Chauviere07, Dickinson94, Dickinson00, Hillen06,  Painter09,Alarcon03,Friedl03} also 
	lead to formation of asymmetrical shape patterns of cell populations.
	The geometrical characterisitcs of animal groups are also of interest, e.g. the impact of anisotropic interactions between individuals
	on the shape of groups have been analyzed recently in  \cite{Cristiani11}.
}

	In this concern, it is worthwile to expose the asphericity $A$ and prolateness $S$ as shape characteristics
 of an {\it instantaneous configuration of a group of particles}, spreading 
in anisotropic environment with preferred orientations.  
{These shape parameters may be of use in mathematical biology: they enable to 
 classify the  shape patterns, formed in processes of bacteria growing in inhomogeneous substrates
or cell invasion in extracellular matrix
of connective tissues with collagen and elastin fibers.
 Since the time-dependent positions of cells or bacteria
in process of migrations  are possible to record in scanning microscopy experiments, these geometrical shape 
characteristics can be directly measurable by analyzing the corresponding image
data by statistical means.}
 Note that recently the parameters $A$ and $S$ were used to describe the instantaneous 
shapes of clouds of particle in turbulent flows in Ref. \cite{Bianchi16}, where it was observed
 an increase of asphericity of such clouds under the action of oriented external fields.
 
In the present work, we exploit the lattice model of BRW spreading 
in anisotropic environment with preferred orientations.  We aim to express the shape characterisitcs $A$ and $S$ in terms of 
fundamental random walk parameters (such as transition probabilities) and analyze how the presence of 
orientational factors and heterogeneity in environment 
  impact the geometrical shape of an instantaneous configuration of group of random walkers. 
 
 As it will be shown, 
such a model allows us in particular, to analyze the spreading process in presence of structural inhomogeneities 
in form of oriented 
lines.  In particular, such a model resembles the arrays of 
oriented nanotubes in  controlled drug delivery  implants, mentioned above.
  
The layout of the paper is as follows. In the next section, we introduce the model and define the observables we are interested in. The analytical expressions
of shape parameters are given in Section \ref{results}, followed by examples  of some model cases presented in Section \ref{examples}.
We end up by giving conclusions and outlook. 

\begin{figure}[t!]
\centering \includegraphics[width=80mm]{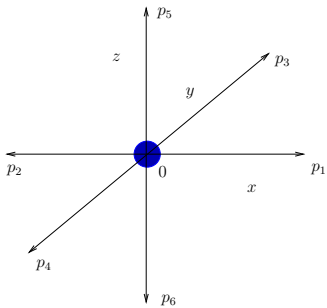}
\caption{ \label{rule}
 Schematic presentation of a random walk process. The particle is putted on a site of a regular $3$-dimensional lattice and 
 at each time step makes a jump towards one of possible $6$ directions with corresponding probability $p_i$ ($i=1,\ldots,6$).
  }
\end{figure}

\section{The model}\label{model}

We start with considering a population of $N$ random walkers, spreading on an infinite $3$-dimensional lattice.
At each time step, each the walker jumps towards one of $6$ nearest neighbor sites 
with corresponding probability $p_i$, $i=1,\ldots,6$ such that $\sum_{i=1}^{6} p_i=1$ (as schematically shown on Fig. \ref{rule}). In the simplest case 
of isotropic uncorrelated random walk, all the transition probabilities are equal: $p_i=\frac{1}{6}$. 
We assume, that all the walkers start to move from the same starting point (let it be the center of coordinate system), so that
we start with highly dense configuration (``drop"), localized in space.
The walkers move independently, without any interactions or correlations between them.  

The probability $P(t,\{n_i^{a}\})$ that the walker $a$  had
performed  $n_i^ {a}$ steps in directions $i$ ($i=1,\ldots,6$) after the total amount of $t$ steps (so that $t=\sum_{i=1}^{6}n_i^ {a}$) is given by:
	\begin{equation}
	P(t,\{n_i^ {a}\})=\frac{t!} {\prod\limits_{i=5}^5n_i^ {a}!\left(t-\sum\limits_{i=1}^5n_i^a\right)!}\, \prod_{i=1}^6p_i^{n_i^ {a}}, \label{Prob}  
	\end{equation}  
	with $\sum_{\{n_i^a \}}P(t,\{n_i^ {a}\})=1$.
	{ 
	
Thus, we can define  the configurational averaging of any observable ${\cal O}$ over all possible trajectories of particle $a$ 
at time $t$ accordingly to:
\begin{equation}
\langle {\cal O} \rangle = {\cal O} P(t,\{n_i^ {a}\}). \label{conf}
\end{equation}
}
To evaluate e.g. the mean value of $\langle n_1^ {a}\rangle$,
we first ``sum out'' the remaining $n_i^{a}$ with $i>1$ in Eq. (\ref{Prob})  according to:
\begin{eqnarray}
&&P(t,n_1^a)=\sum_{n_2^a=0}^{t-n_1^a}\,\,\ldots\sum_{n_5^{a}=0}^{t-n_1^a-n_2^a-n_3^a-n_4^a} P(t,\{n_i^ {a}\})=\nonumber\\
&&=\frac{t!}{n_1^ {a}!(t-n_1^a)!}p_1^{n_1^a}(1-p_1)^{t-n_1^a}, \label{sumout}
\end{eqnarray}
so that:
	\begin{eqnarray}
&&\langle n_1^ {a}\rangle=\sum_{n_1^ {a}=0}^t n_1^ {a}P(t,n_1^a) =tp_1,
\end{eqnarray}
 and in general:
	\begin{eqnarray}
&&\langle n_i^ {a}\rangle=\sum_{n_i^ {a}=0}^t n_i^ {a} P(t,n_i^ {a})=tp_i,
\\
&&\langle (n_i^ {a})^2\rangle=\sum_{n_i^ {a}=0}^t (n_i^ {a})^2 P(t,n_i^ {a})= tp_i+t(t-1)p_i^2, \label{n2} \\
&&\langle n_i^ {a} n_j^ {a}\rangle=\sum_{n_i^ {a}=0}^t \sum_{n_j^ {a}=0}^{t-n_i^ {a}} n_i^ {a}n_j^ {a} P(t,n_i^ {a},n_j^a)
 = t(t-1)p_ip_j,\,\,i\neq j,
\end{eqnarray}
in the last equation, $P(t,n_i^ {a},n_j^a)$ is obtained by summing out all $n_k^a$ with $k\neq i$ and $k\neq j$ in the same way as in Eq. (\ref{sumout}). 

Since we assume,
that walkers $a$ and $b$ move independently, we also have:
\begin{equation}
\langle n_i^{a} n_j^{b}\rangle=\sum_{n_i^ {a}=0}^t n_i^ {a} P(t,n_i^{a}) \sum_{n_j^ {b}=0}^{t} n_j^ {b} P(t,n_j^ {b})
 = t^2p_ip_j.
\end{equation}

\section{Results: Universal shape parameters}\label{results}

Let $\vec{R}_a(t)=\{x^a(t),y^a(t),z^a(t)\}\equiv\{x_{1}^a(t),x_{2}^a(t),x_{3}^a(t)\}$ be the position vector of the $a$th walker at time $t$ ($a=1,\ldots,N$).
The position of each walker, averaged over an ensemble of all possible trajectories of particle, can be easily obtained
using the results of previous Section. Really, since e.g. the averaged coordinate $\langle x_1(t) \rangle$ of
 a walker is given by a difference of number of steps to the right and to the left along the $x_1$-axis, we have:  
	\begin{eqnarray}
	&& \langle x_1^a(t) \rangle  = \langle n_1^a-n_2^a \rangle = t(p_1-p_2),\nonumber\\
&&	\langle x_2^a(t) \rangle = \langle n_3^a-n_4^a \rangle = t(p_3-p_4), \nonumber \\
&& 	\langle x_3^a(t) \rangle = \langle n_5^a-n_6^a \rangle = t(p_5-p_6),
\end{eqnarray}
so that: $\langle x_i^a(t) \rangle   = t(p_{2i-1}-p_{2i})$. 


Correspondingly:
\begin{eqnarray}
&&	\langle (x_i^a(t))^2 \rangle  =  t(p_{2i-1}+p_{2i})+t(t-1)(p_{2i-1}-p_{2i})^2,\label{cor1}\\
&&	 \langle x_i^a(t)x_j^a(t) \rangle  =t(t-1)(p_{2i-1}-p_{2i})(p_{2j-1}-p_{2j}),\,\,i\neq j \\
	&&	\langle x_i^{a}(t) x_i^{b}(t)\rangle =  t^2(p_{2i-1}-p_{2i})^2, \\ 
	&&	\langle x_i^{a}(t) x_j^{b}(t)\rangle =  t^2(p_{2i-1}-p_{2i})(p_{2j-1}-p_{2j}).\label{cor2}   
		      \end{eqnarray}

The shape properties of configuration of the population can be characterized \cite{Aronovitz86, Rudnick86}
in terms of the gyration tensor $\bf{Q}$ with components:
\begin{equation}
\langle Q_{ij}(t) \rangle =\frac{1}{N(N-1)}\underset{a<b}{\sum_{a,b=1}^N} \langle (x_{i}^a(t)-x_{i}^b(t))(x_{j}^a(t)-x_{j}^b(t))\rangle,\,\,\,\,\,i,j=1,2,3.
\label{mom}
\end{equation}
The spread in eigenvalues $\lambda_i(t)$ $(i=1,2,3)$ of the gyration tensor describes the distribution of particles in configuration and
thus measures the asymmetry of a shape. In particular, in completely isotropic symmetric case 
 all the eigenvalues $\lambda_i(t)$ are equal.

Let us introduce  the rotationally invariant
universal combinations of components of the gyration tensor \cite{Aronovitz86, Rudnick86}.
Let ${{\lambda}}_{{\rm av}}(t)\equiv {\rm Tr}\, {\bf{Q}}/3$
be the average eigenvalue of  the gyration tensor.
Then the extent of asphericity of an instantaneous configuration of population 
is characterized by quantity $A$ defined as:
\begin{eqnarray}
&& A =\frac{1}{6} \sum_{i=1}^3\frac{(\lambda_{i}(t)-{{\lambda_{{\rm av}}(t)}})^2}{{\lambda_{{\rm av}}(t)}^2}= 
\Big( \sum_{i=1}^3\langle Q_{ii}(t)\rangle^2+ 3 \,\,\underset{i<j}{\sum\limits_{i,j=1}^3} \langle Q_{ij}(t)\rangle^2\nonumber
\\
&&- \underset{i<j}{\sum\limits_{i,j=1}^3}\langle Q_{ii}(t)\rangle \langle Q_{jj}(t)\rangle  \Big)/\Big( \sum_{i=1}^3\langle Q_{ii}(t)\rangle  \Big)^2.\label{add} 
\end{eqnarray}
This universal quantity equals zero for a completely isotropic spherical configuration, where all the
 eigenvalues are equal $\lambda_i=\overline{\lambda}$, and takes a maximum value
of one in the case of a stretched highly anisotropic configuration, where all the eigenvalues equal zero except of one.
Thus, the inequality holds: $0\leq A\leq 1$.
Another rotationally invariant quantity, defined in three dimensions, is the so-called prolateness $S$:
\begin{eqnarray}
&&S =\frac{\prod_{i=1}^3(\lambda_{i}(t)-{{\lambda_{{\rm av}}(t)}})}{{{\lambda_{{\rm av}}(t)}}^3}=\Big( 2  \sum_{i=1}^3\langle Q_{ii}(t)\rangle^3
-3\underset{i\neq j}{\sum\limits_{i,j=1}^3}\langle Q_{ii}(t)\rangle^2 \langle Q_{jj}(t) \rangle\nonumber \\
&&-18\underset{i\neq j \neq k}{\sum_{i,j,k=1}^3}\langle Q_{ii}(t)\rangle \langle Q_{jk}(t) \rangle^2+
9\underset{i\neq j}{\sum_{i,j=1}^3} \langle Q_{ii}(t)\rangle \langle Q_{ij}(t) \rangle^2\nonumber\\
	&&+12\prod_{i=1}^3\langle Q_{ii}(t)\rangle +54\langle Q_{12}(t)\rangle\langle Q_{13}(t)\rangle\langle Q_{23}(t)\rangle 
	\Big)/\Big( \sum_{i=1}^3\langle Q_{ii}(t)\rangle  \Big)^3.  \label{sdd}
\end{eqnarray}
For absolutely prolate, stretched rod-like configuration ($\lambda_1\neq0,\lambda_2=\lambda_3=0$), the parameter $S$ equals two,
whereas for absolutely oblate, disk-like shape  ($\lambda_1=\lambda_2,\lambda_3=0$) it takes on a value of $-1/4$.
In general, $S$ is positive for prolate ellipsoid-like shape  ($\lambda_1\gg \lambda_2\approx\lambda_3$) and negative for oblate
ones ($\lambda_1\approx\lambda_2\gg\lambda_3$), whereas its magnitude measures how oblate or prolate the configuration is.

Next, we will find the exact values of quantities (\ref{add}) and (\ref{sdd}) for a system of non-interacting $N$ asymmetric 
random walkers on a lattice.

Expressions (\ref{cor1})-(\ref{cor2}) allow us to find  the averaged components of gyration tensor.	For example:	      
	\begin{eqnarray}
	&& {\langle Q_{11}(t) \rangle} = \frac{1}{N(N-1)} \left \langle \underset{a<b}{\sum_{a,b=1}^N }(x_1^{a}(t)-x_1^{b}(t))^2 \right \rangle  \nonumber\\
	&&=\langle (x_1^a(t))^2 \rangle -  \langle x_1^a (t)x_1^b(t)\rangle = t\left((p_1+p_2)-(p_1-p_2)^2\right).\label{Qxx}
	\end{eqnarray}
	So in general:
	\begin{eqnarray}
	&& {\langle Q_{ii}(t) \rangle} =  t\left((p_{2i-1}+p_{2i})-(p_{2i-1}-p_{2i})^2\right),\label{Qii}\\
	&& {\langle Q_{ij}(t) \rangle} = -t(p_{2i-1}-p_{2i})(p_{2j-1}-p_{2j}).\label{Qij}
	\end{eqnarray}

	Finally, substituting the values (\ref{Qii}),  (\ref{Qij}) into Eqs. (\ref{add}) and (\ref{sdd}), we receive expressions for shape parameters $A$ and $S$ of the system:
	\begin{eqnarray}
	&&A(\{p_i\})=\Big(\sum_{i=1}^3 (p_{2i-1}+p_{2i}-(p_{2i-1}-p_{2i})^2)^2
	\nonumber  \\
	&& -\underset{i<j}{\sum\limits_{i,j=1}^3}\Big[ (p_{2i-1}+p_{2i}-(p_{2i-1}-p_{2i})^2) (p_{2j-1}+p_{2j}-(p_{2j-1}-p_{2j})^2)
	\nonumber \\
	&& +3(p_{2i-1}-p_{2i})^2(p_{2j-1}-p_{2j})^2\Big]
	 \Big)/\Big(\sum_{i=1}^6p_i-\sum_{i=1}^3(p_{2i}-p_{2i-1})^2\Big)^2,\label{afinal}
	 \end{eqnarray}
	 \begin{eqnarray}
	 && S(\{p_i\})=\left(12\prod_{i=1}^3\left((p_{2i-1}+p_{2i}-(p_{2i-1}-p_{2i})^2)\right. \right. \nonumber\\
	 && +2 \sum_{i=1}^3 (p_{2i-1}+p_{2i}-(p_{2i-1}-p_{2i})^2)^3  \nonumber\\
	 &&-3\left. \underset{i\neq j}{\sum\limits_{i,j=1}^3} (p_{2i-1}+p_{2i}-(p_{2i-1}-p_{2i})^2)^2 (p_{2j-1}+p_{2j}-(p_{2j-1}-p_{2j})^2)  \nonumber \right.\\
	&& - 9\underset{i \neq j}{\sum\limits_{i,j=1}^3}(p_{2i-1}+p_{2i}-(p_{2i-1}-p_{2i})^2))(p_{2i-1}-p_{2i})^2(p_{2j-1}-p_{2j})^2\nonumber\\
	&& -18 \underset{i\neq j \neq k}{\sum_{i,j,k=1}^3}(p_{2i-1}+p_{2i}-(p_{2i-1}-p_{2i})^2))(p_{2j-1}-p_{2j})^2(p_{2k-1}-p_{2k})^2\nonumber\\
 &&\left.-54\prod_{i=1}^3(p_{2i-1}-p_{2i})^2
	 \right)/\left(\sum_{i=1}^6p_i-\sum_{i=1}^3(p_{2i-1}-p_{2i})^2\right)^3. \label{sfinal}
	\end{eqnarray}
The quantities (\ref{afinal}), (\ref{sfinal}) are universal in the sence 
that they do not depend either on time or on number of particles, 
and appear to be the  functions only of transition probalities $p_i$. Note, that in real experiments and in computer simulations, 
one would obtain the quantities given by (\ref{afinal}), (\ref{sfinal}) as the asymptotical values in the limits of large 
enough number of particles and long enough time of spreading process ($t\to\infty$, $
N\to\infty$).

 {
  
The above scheme can be generalized to the case, when one has random values $p_{i}(\vec{r})$ 
(with $\vec{r}=\{x,y,z \}$) at different sites of the lattice, taken from some
 distribution $\rho(\{p_{i}(\vec{r})\})$, so that the averaged 
 value 
 ${\overline{p_{i}}}$ is given by
 \begin{equation}
 {\overline{p_{i}}}=\sum p_{i}(\vec{r})\rho(\{p_{i}(\vec{r})\}). 
 \end{equation} 
Such a situation may occur due to the action of local random fields or presence of structural defects, occupying some sites of the lattice. 
Dealing with such systems that display randomness of structure,
one should perform the double averaging for all quantities of interest \cite{Braut59,Emery75}: first, the 
 configurational averaging over all possible trajectories of particles according to (\ref{conf}),
  and then the averaging
 over all different realizations of disorder according to:
 \begin{equation}
 {\overline{ \langle {\cal O} \rangle} }=\sum\rho(\{p_{i}(\vec{r})\}) \langle {\cal O} \rangle= {\cal O} {\overline {P(t,\{n_i^ {a}\})}}.  \label{av}
 \end{equation} 
 Here, ${\overline {{P(t,\{n_i^{a}\})}} }$ is the distribution function (\ref{Prob}) averaged over disorder realizations of disorder:
	\begin{equation}
{\overline{	P(t,\{n_i^ {a}\})}}=\frac{t!} {\prod\limits_{i=1}^5 n_i^ {a}!\left(t-\sum\limits_{i=1}^5n_i^a\right)!}  
  \sum \rho(\{p_{i}(\vec{r})\}) \prod_{i=1}^6 \prod_{k=1}^{n_i^a} p_{i}(\vec{r}_k) \label{ppdis}  
	\end{equation} 
	where $p_{i}(\vec{r}_k)$ denotes a probability to jump in direction $i$ for a walker which is located at 
	site $\vec{r}_k$ of a lattice.       
	In the case, when there are no correlations in distribution of $p_{i}(\vec{r})$ at different sites of the lattice, we can rewrite in last expression:
	\begin{equation}
	\sum \rho(\{p_{i}(\vec{r})\}) \prod_{k=1}^{n_i^a} p_{i}(\vec{r}_k) =\prod_{k=1}^{n_i^a} \sum \rho(\{p_{i}(\vec{r})\}) p_{i}(\vec{r}) ={\overline p_{i} }^{n_i^a},  
	\end{equation}  
  so that all of the observables of interest in relations above 
  can be obtained just by substituting $p_i$ by ${\overline p_{i} } $  
 in corresponding equations.
 }
  
In the next section, we illustrate the results obtained by considering several model cases of invasion in anisotropic environment.

	\begin{figure}[t!]
\centering\includegraphics[width=120mm]{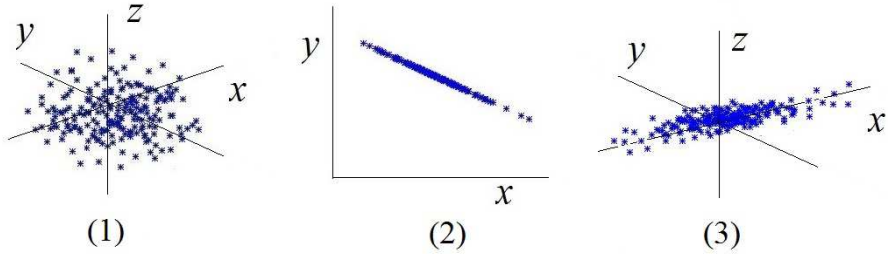}
\caption{ \label{exam}
Instantaneous  configurations of population of $N=100$ random walkers on a lattice after $t=2000$ steps, results of computer simulations. (1) 
 Isotropic case with $p_i=1/6$. The values of shape parameters $A=0$, $S=0$.
 (2) Moving on the half-space of $xy$ plane with $p_1=0.4$, $p_3=0.6$, $p_2=p_4=p_5=p_6=0$. 
 Maximal anisotropic state with  $A=1$, $S=2$. 
  (3) The case, when moving in $x$ direction is more probable, than in others: $p_1=p_2=0.4$,
 $ p_3=p_4=p_5=p_6=0.05$. The values of shape parameters: $A=0.49$, $S=0.68$.
  }
\end{figure}

\section{Examples}\label{examples}

1) In the most trivial isotropic case, when all $p_i$ are equal ($p_i=1/6$), we have: $A=0$, $S=0$ (see Fig. \ref{exam}(1)).

2) Let us consider the case, when $p_3=1-p_1$, $p_2=p_4=p_5=p_6=0$: the population is moving on the half-space of $xy$ plane
with non equal transition probabilities in $x$ and $y$ directions.
It appears, that independently on the  $p_1$ and $p_3$ values, we receive highly anisotropic, completely stretched configuration with $A=1$, $S=2$ (Fig. \ref{exam}(2)). 

3) Next, let us assume $p_1=p_2$, $p_3=p_4=p_5=p_6=(1-2p1)/4$: moving along the $x$-axis is more (or less) probable, then in two 
remaining directions	 (Fig. \ref{exam}(3)). In this case, on the basis of (\ref{afinal}), (\ref{sfinal}) we obtain:
\begin{eqnarray}                                         
               && A(p_1)= -3p_1 + 9p_1^2  + \frac{1}{4}, \label{aap} \\
                 && S(p_1)=\frac{1}{4}(6p_1-1)^3.  \label{ssp}                         
\end{eqnarray}

\begin{figure}[b!]
\centering \includegraphics[width=70mm]{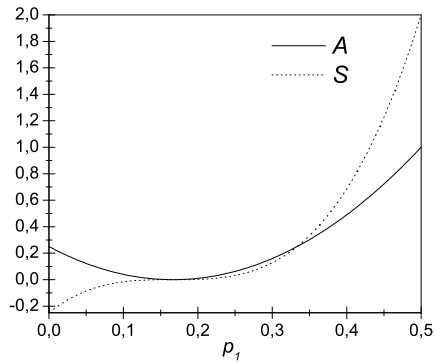}
\caption{ \label{aa}
  Shape parameters $A$ (as given by Eq. (\ref{aap})) and $S$ (given by (\ref{ssp})) as functions of probability $p_1=p_2$ 
  (whereas  $p_3=p_4=p_5=p_6=0$). 
   }
\end{figure}

Parameters $A$ and $S$ as functions of $p_1$ are shown on Fig. \ref{aa}. Note, that $p_1$ can vary in this case from 0 to $1/2$. 
At $p_1=\frac{1}{6}$, we restore the isotropic case (1) with $A=S=0$. Further increasing of $p_1$ 
leads to growing of anisotropy, until it reaches the final stage (configuration completely stretched along $x$ axis)
 at $p_1=p_2=\frac{1}{2}$,  $p_3=p_4=p_5=p_6=0$.

4)   Finally, let us consider the most interesting case, when population of particles is spreading in environment with 
structural inhomogeneities (obstacles)
in the form of parallel lines, randomly distributed in $xy$ plane and oriented along $z$ axis (see Fig. \ref{lines}). 
Let $c$ be the concentration of lines ($0\leq c \leq 1$). From the point of view of each random walker, presence of lines does not 
prevent  jumps in $z$ direction, but plays an essential role for movement in $xy$ plane. Really, for each lattice site, 
one of 4 nearest neighbors in $xy$ plane can be occupied with probability $c$ (belonging to the line) and is thus not allowed 
for random walker. The probability, that $k$ nearest neighbors ($0\leq k \leq 4$) are occupied, is given by Bernoulli formula for 
binomial probability distribution:
\begin{equation}
\rho(k)=\frac{4!}{k!(4-k)!}c^k(1-c)^{4-k}. \label{ber}
\end{equation} 

\begin{figure}[t!]
\centering \includegraphics[width=70mm]{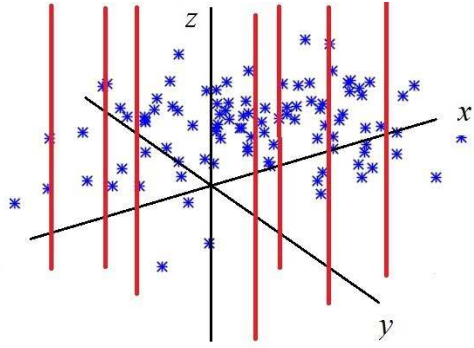}
\caption{ \label{lines}
 Schematic presentation of an instantaneous configuration of population spreading on a 
 lattice in the presence of
 array of lines (tubes) oriented along $z$ axis.  
  }
\end{figure}

 \begin{figure}[b!]
\centering \includegraphics[width=70mm]{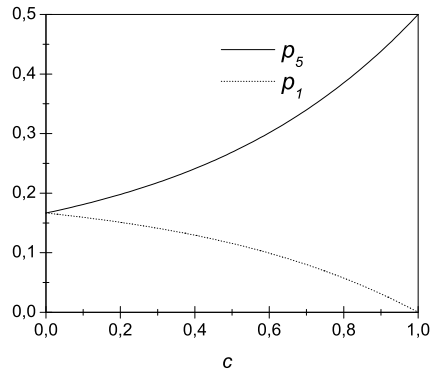}
\caption{ \label{pp}
 Averaged transition probabilities of moving in $z$ direction (as given by Eq. (\ref{pz})) and in $xy$ plane  (Eq. (\ref{px})
  as functions of concentration $c$ of oriented lines.  }
\end{figure}

The given problem thus essentially differs from examples above, where transition probabilities were
spatially-independent constants $p_i$. 
Here, due to randomness of defects distribution in space, at each lattice site 
we observe  one of possible $p_{ik}=1/(6-k)$  with corresponding probabilities $\rho(k)$ given by (\ref{ber}).
Namely, for $i=5,6$ (transition probabilities in $z$ direction) taking into account Eq. (\ref{av}) we have: 
\begin{eqnarray}
&&{\overline{p_i}}=\sum_{k=0}^4p_{ik}\rho(k)=\sum_{k=0}^4\frac{1}{6-k}\rho(k)=\frac{1}{6}(1-c)^4+\frac{4}{5}c(1-c)^3\nonumber\\
&&+\frac{3}{2}c^2(1-c)^2+\frac{4}{3}c^3(1-c)+\frac{1}{2}c^4,\,\,\,\,\,i=5,6.\label{pz}
\end{eqnarray}
The corresponding transition probabilities $p_{ik}$ in $xy$ planes are smaller 
by the factor $(4-k)/4$ due to the fact, that $k$ jumps are forbidden by presence of defects, so that:   
\begin{eqnarray}
&&{\overline{p_i}}=\sum_{k=0}^4\frac{4-k}{4}\frac{1}{6-k}\rho(k)=\frac{1}{6}(1-c)^4+\frac{3}{5}c(1-c)^3\nonumber\\
&&+\frac{3}{4}c^2(1-c)^2+\frac{1}{3}c^3(1-c),\,\,\,\,\,i=1,\ldots,4. \label{px}
\end{eqnarray}
As expected, we have:
\begin{equation}
\sum_{i=1}^6{\overline{p_{i}}}=1.
\end{equation}
The functions (\ref{pz}) and (\ref{px}) are presented graphically on Fig. \ref{pp}. 
At $c=0$, we restore the pure isotropic case when all ${\overline{p_i}}=1/6$. Increasing of $c$ leads to separation of these quantities: transition probabilities 
of moving in $z$ direction are growing, whereas corresponding values in $xy$ plane are gradually tending to zero. Approaching the very large values of 
$c$ around 1, when there are practically no possibility of moving in $xy$ plane, ${\overline{p_5}}$ and ${\overline{p_6}}$ reach the limiting values of $1/2$.
 Note also, that $c=1$ (fully occupied lattice) has no physical meaning in our case, since there is no possibility for particles to move at all.

{
 
Let us recall, that in problem under consideration we need to 
perform the averaging of function (\ref{ppdis}) over distribution of $p_{i}(\vec{r})$. Note, that since the structural defects are randomly
distributed in $xy$ plane, there is in principle no correlations in $p_{i}(\vec{r})$ and $p_{j}(\vec{r'})$ with $i,j=1,\ldots,4$ 
at different sites of the lattice. 
However, since defects have a form of lines oriented along $z$ axis, the correlations occur 
in corresponding $p_{i}(x,y,z)$ and $p_{j}(x,y,z\pm 1)$ (namely, always $p_{i}(x,y,z)=p_{j}(x,y,z\pm 1)$ with $i,j=5,6$). So, 
when the walking particle is performing a series of $l$ consequent steps 
in direction $z$, we have in (\ref{ppdis}): 
\begin{eqnarray}
&&\sum_{k=0}^4 \rho(k) \prod_{m=1}^l p_{i}(\vec{r}_m)=\sum_{k=0}^4 \rho(k) p_{ik}^l=\sum_{k=0}^4 \rho(k) \left( \frac{1}{6-k} \right)^l=(1-c)^4\left(\frac{1}{6} \right)^l\nonumber\\
&&+4c(1-c)^3\left(\frac{1}{5} \right)^l+6c^2(1-c)^2\left(\frac{1}{4} \right)^l+4c^3(1-c)\left(\frac{1}{3} \right)^l
+c^4\left(\frac{1}{2} \right)^l, \label{corz} 
 \end{eqnarray}
whereas in case of no correlations between $p_{i}(\vec{r})$ one would have:
\begin{eqnarray}
\sum_{k=0}^4 \rho(k) \prod_{m=1}^l p_{i}(\vec{r}_m)=\prod_{m=1}^l \sum_{k=0}^4 \rho(k) p_{ik}=\overline{p_i}^l,             
\label{uncorz}
\end{eqnarray}
with $\overline{p_i}$ given by Eq. (\ref{pz}). Taking into account, that $c$ is a small parameter, and comparing expressions
(\ref{corz}) and (\ref{uncorz}) at fixed values of $c$ and $l$, we found that difference between them is of order of magnitude $10^{-3}$ at $l=2$ 
but becomes negligibly small at large $l$. To avoid difficulties with taking into account all possibilities of series of subsequent steps in $z$ direction,
in what follows we will make use of assumption  
(\ref{uncorz}), leading to simple substituting $p_i$ by ${\overline p_{i}} $  
 in final expressions for observables.
  We stress however, that due to this assumption our results are rather of qualitative character and are reliable at
  small values of concentration $c$.   
}
\begin{figure}[t!]
\centering \includegraphics[width=70mm]{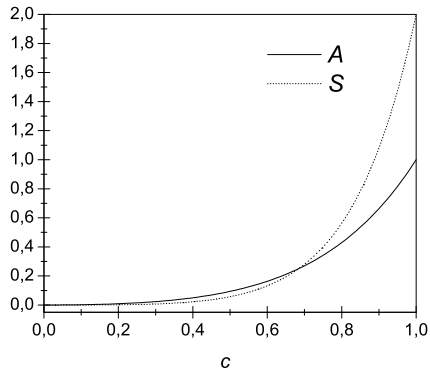}
\caption{ \label{aalines}
  Shape parameters $A$ (given by Eq. (\ref{aline})) and $S$ (Eq. (\ref{sline})) as functions of concentration $c$ of oriented lines.  }
\end{figure}

Substituting  (\ref{pz}), (\ref{px}) into expressions for the shape parameters (\ref{afinal}), (\ref{sfinal}), we find corresponding expressions:
\begin{eqnarray}
&&A = \frac{1}{100}c^2(c^6+4c^5+10c^4+20c^3+25c^2+24c+16),\label{aline}\\
&&S=\frac{1}{500}c^3(c^9+6c^8+21c^7+56c^6+111c^5+174c^4\nonumber\\
&&+219c^3+204c^2+144c+64).\label{sline}
\end{eqnarray}
Expressions (\ref{aline}) and (\ref{sline}) as functions of $c$ are plotted on Fig. \ref{aalines}. 
As expected, the asymmetry of shape is increasing with growing of the concentration of structural defects:
the configuration of spreading population becomes more and more elongated in $z$ direction. Thus, the presence of an array of oriented fibers (tubes) leads
to considerable spatial organization of invading population in environment.

\section{Conclusions}

In the present work, we developed the simple mathematical model of population of agents, spreading in heterogeneous 
environment which is characterized by spatial anisotropy. 
It can be related to numerous processes, encountered in biology and medicine, such as chemotaxis or gravitaxis,
the cell invasion in extracellular environment,  migrations of living
 organisms in oriented habitats with non-homogeneous spatial distributions of resources etc. 

 The presence of orientational factors in environment induces significant changes in  
geometrical shape of instantaneous configuration and spatial organization of spreading population. 
The motivation of the present study was to analyze the asphericity $A$ and prolateness $S$ as shape characteristics
 of instantaneous configuration of a group of particles, spreading 
in anisotropic environment with preferred orientations. 
  For this purpose, we studied a system of non-interacting $N$ random walkers, spreading on 
 an infinite $d=3$-dimensional lattice, starting with initial very dense configuration (``drop'').
 The transition probabilities $p_i$ ($i=1,\ldots,6$) in different directions are assumed to be non-equal (asymmetric).
   We expressed the shape characterisitcs in terms of 
fundamental random walk parameters  and  found the exact analytical values for 
the parameters,  such as asphericity $A$ (Eq. \ref{afinal}) 
  and prolateness $S$ (Eq. \ref{sfinal}) as functions of transition probabilities $p_i$.
 {These ideas can find their application in mathematical methods of analysis and classification of asymmetric shape pattern 
 	formation is processes of bacteria colonies dynamics in inhomogeneous substrates or in presence of
 	gradient of a chemotactic factors,  as well as
 	 cell migration in  tissue matrices. }

To illustrate the results obtained, we considered several model cases of oriented environment. Of particular interest 
is the case, when the array of lines (tubes)
 of parallel orientation is present in the system. This can serve as a model of extracellular matrix
with collagen and elastin fibers or systems of oriented nanotube array in drug delivery implants.
It is quantitatively shown, that presence of such objects leads
to considerable spatial orientation and organization of invading population.


\begin{thebibliography}{90}

\bibitem{Alarcon03}
T. Alarc\'on,  H.M. Byre, and P.K. Main,  A cellular automaton model for tumour growth in
inhomogeneous environment. {\it J. Theor. Biol.} { \bf 225} (2003)  257-274.


\bibitem{Alt80}
W. Alt,   Biased random walk model for chemotaxis and related diffusion approximation. {\it J.
Math. Biol.} { \bf 9} (1980) 147-177.

\bibitem{Anderson00}
A.R.A. Anderson, M.A.J. Chaplain, E.L. Newman, R.J.C. Steele, and A.M. Thompson,  Mathematical
modelling of tumour invasion and metastasis. {\it J. Theor. Med.} { \bf 2} (2000) 129-154.

\bibitem{Armsworth99}
P.R. Armsworth and  L. Bode,  The consequences of non-passive advection and directed motion for population
dynamics. {\it Proc R Soc Lond A} { \bf 455} (1999) 4045-4060. 

\bibitem{Armsworth05}
P.R.  Armsworth and  J.E. Roughgarden,    The impact of directed versus random movement on population
dynamics and biodiversity patterns.  {\it Am. Nat.} { \bf 165} (2005) 449-465

\bibitem{Aronovitz86}
 J.A. Aronovitz and D.R. Nelson,  Universal features of polymer shapes. {\it  J. Physique} { \bf 47} (1986) 1445-1456.



\bibitem{Belgacem95}
F. Belgacem and  C. Cosner,    The effects of dispersal along environmental gradients on the dynamics of populations in
heterogeneous environments. {\it  Can. Appl. Math. Quar.} { \bf 3} (1995) 379-397.

\bibitem{Jacob04}
E. Ben-Jacob, H. Levine,   The artistry of nature.  {\it Nature}. {\bf 409} (2001) 985?986


\bibitem{Berg00}
H.C. Berg,  Motile Behavior of Bacteria. {\it  Physics Today} { \bf 53}  (2000) 24-29.

\bibitem{Bianchi16}
 S. Bianchi, L.  Biferale,  A. Celani, and M.  Cencini,  On the evolution of particle-puffs in turbulence.
  {\it   Eur. J. Mech. B Fluids} { \bf 55} (2016)  324-329. 

\bibitem{Bishop88}
M. Bishop and C.J. Saltiel,  Polymer shapes in two, four, and five dimensions. {\it  J. Chem. Phys.}
 {\bf 88} (1988) 3976-3980.


\bibitem{Braut59}
R. Brout,  Statistical Mechanical Theory of a Random Ferromagnetic System. {\it  Phys. Rev.} {\bf 115} (1959) 824-835.

\bibitem{Emery75}
V.J. Emery,  Critical properties of many-component systems. {\it  Phys. Rev. B } {\bf 11} (1975) 239-247. 

\bibitem{Cantrell06}
R.S.  Cantrell, C. Cosner, and Y. Lou,   Movement toward better environments and the evolution of rapid
diffusion, {\it  Math. Biosci.} { \bf 204} (2006) 199-214. 


\bibitem{Chalub04}
F.A.C.C.  Chalub, P.A. Markowich, B. Perthame, and C. Schmeiser,  Kinetics models for
chemotaxis and their drift-diffusion limits. {\it  Monatsh. Math.} {\bf  142} (2014) 123-141.

\bibitem{Chauviere07}
A. Chauviere, T. Hillen, and L. Preziosi,   Modeling cell movement in anisotropic and heterogeneous
network tissues. {\it  Networks and Heterogeneous Media} { \bf 2} (2007) 333-357.

\bibitem{Codling17}
E.A. Codling,  M.J Plank, and S. Benhamou,  Random walk models in biology. {\it  J. R. Soc. Interface}  { \bf 5} (2008) 813-834.

\bibitem{Cristiani11}
E. Cristiani, P. Frasca, and B. Piccoli,  Effects of anisotropic interactions on the structure
of animal groups. {\it  J. Math. Biol.}  {\bf 62} (2011) 569-588.

\bibitem{Dewhirst09}
S. Dewhirst and  F. Lutscher,   Dispersal in heterogeneous habitats: thresholds, spatial scales, and approximate
rates of spread. {\it  Ecology} { \bf 90} (2009) 1338-1345.

\bibitem{Dickinson94}
R.B.  Dickinson, S. Guido, and R.T. Tranquillo,   Biased cell migration of fibroblasts exhibiting
contact guidance in oriented collagen gels. {\it  Ann. Biomed. Eng.} { \bf 22} (1994) 342-356.

\bibitem{Dickinson00}
R.B. Dickinson  A generalized transport model for biased cell migration in an anisotropic
environment. {\it  J. Math. Biol.} { \bf 40}  (2000) 97-135.

\bibitem{Diehl89}
H.W. Diehl  and E. Eisenriegler,  Universal shape ratios for open and closed random walks:
exact results for all $d$. {\it  J. Phys. A: Math. Gen.} {\bf 22} (1989)  L87-L91.


\bibitem{Dolak05}
Y.  Dolak  and  C. Schmeiser,  Kinetic models for chemotaxis: Hydrodynamic limits and spatiotemporal
mechanics. {\it   J. Math. Biol.} { \bf 51} (2005)  595-615.


\bibitem{Franco02}
C. Di Franco, E. Beccari, T. Santini, G. Pisaneschi, and G. Tecce, Colony shape as a genetic trait in the pattern-forming Bacillus mycoides,
{\it BMC Microbiol.} {\bf 2} (2002) 33(1-15).

\bibitem{Friedl03}
 P. Friedl and K. Wolf,    Tumour-cell invasion and migration: diversity and escape mechanisms.
{\it  Nature Rev.} {\bf 3} (2003) 362-374.

\bibitem{Gaspari87}
G. Gaspari, J. Rudnick, and A Beldjenna,   The shapes of open and closed random walks: a $l/d$ expansion,
 {\it  J. Phys. A: Math. Gen.} {\bf 20} (1987)  3393-3414.
 
\bibitem{Hill97}
N.A. Hill and D.P.  H\"ader,     A biased random walk model
for the trajectories of swimming micro-organisms.
{\it  J. Theor. Biol.} { \bf 186} (1997) 503-526.

\bibitem{Hillen06}
T. Hillen,  M$^5$ mesoscopic and macroscopic models for mesenchymal motion. {\it  J. Math. Biol.}
{\bf 53} (2006) 585-616.

\bibitem{Hillen13}
T. Hillen and K.J.  Painter,  Transport and Anisotropic Diffusion Models for Movement
in Oriented Habitat. In:  Dispersal, Individual Movement and Spatial Ecology. Lecture Notes in Mathematics, vol 2071 (2013) Springer,  Berlin,
pp 177-222.


\bibitem{Kirk97}
R.W. van Kirk and  M.A. Lewis,  Integrodifference models for persistence in fragmented habitats.
{\it  Bull. Math. Biol.} {\bf 59} (1997) 107-137.

\bibitem{Kuhn34}
W. Kuhn,  \"Uber die Gestalt fadenf\"ormiger Molek\"ule in L\"osungen, {\it  Kolloid-Z.} {\bf 68} (1934)  2-15.

\bibitem{Losic15}
D. Losic, M. Aw, A. Santos, K.  Gulati, and M. 
Bariana,  Titania nanotube arrays for local
drug delivery: recent advances
and perspectives. {\it  Expert Opin. Drug Deliv.}  { \bf 12} (2015) 103-127.

\bibitem{Mendelson76}
NH Mendelson, Helical growth of Bacillus subtilis: a new model of cell growth. {\it Proc Natl Acad Sci USA} (1976) {\bf 73} (1976) 1740-1744

\bibitem{Mokhtari13}
Z. Mokhtari et al.,  Automated Characterization and Parameter-Free
Classification of Cell Tracks Based on Local Migration
Behavior. {\it  PLoS One} {\bf 8} (2013) e80808(1-20).


\bibitem{Pedley90}
T.J. Pedley and J.O.  Kessler,  A new continuum model for
suspensions of gyrotactic micro-organisms. {\it  J. Fluid Mech.} {\bf 212} (1990)
155-182.


\bibitem{Painter09}
K.J. Painter,  Modelling cell migration strategies in the extracellular matrix. 
{\it  J. Math. Biol.} {  \bf 58} (2009) 511-
543. 

\bibitem{Patlack53}
C.S. Patlack,  Random walk with persistence and external bias. 
{\it  Bull. Math. Biophys.} {\bf 15}  (1953) 311-338.

\bibitem{Rudner98}
R. Rudner, O. Martsinkevich, W. Leung, and E.D. Jarvis ED,
 Classification and genetic characterization of pattern forming Bacilli. {\it Mol. Microbiol.} {\bf 27} (1998) 687-703.

\bibitem{Rudnick86}
 J. Rudnick and G.  Gaspari,  The aspherity of random walks. 
 {\it   J. Phys. A } { \bf 19} (1986) L191-L194.


\bibitem{RWbook}
M.F.  Shlesinger  and B. West B (eds),   Random Walks and their Applications in the Physical and Biological Sciences. AIP Conference Proceedings, vol  109 (1984)  AIP,  New York
 
 \bibitem{Salhi93}
 B. Salhi and N. Mendelson, Patterns of gene expression in
 Bacillus subtilis colonies {\it J. Bacteriol.}  {\bf 175} (1993) 5000-5008
 
 
 \bibitem{Sarris12}
M. Sarris  et al., 
Inammatory chemokines direct and restrict leukocyte migration within live
tissues as glycan-bound gradients. {\it  Current Biology} {\bf 22} (2012) 2375-2382.
 
 
 
 \bibitem{Sciutto94}
S.J. Sciutto,  Study of the shape of random walks. {\it   J. Phys. A Math. Gen.} {\bf 27} (1994) 7015-7034.
 
 
 \bibitem{Shapiro84}
  J. A. Shapiro,  The use of Mudlac transposons as tools for vital
 staining to visualize clonal and non-clonal patterns of organization in bacterial
 growth on agar surfaces. {\it J. Gen. Microbiol.} {\bf  130} (1984) 1169-1181. 
 
 \bibitem{Shapiro85}
  J. A. Shapiro, Scanning electron microscope study of Pseudomonas
  putida colonies. {\it J. Bacteriol.} {\bf  164} (1985) 1171-1181. 
 
 \bibitem{Shapiro95}
 J.A. Shapiro, The significance of bacterial colony patterns. {\it  Bio Essays} {\bf 17} (1995) 597-607.
 
 \bibitem{Solc71}
 K. \u Solc and W.H. Stockmayer,  Shape of a Random-Flight Chain. {\it  J. Chem. Phys.} {\bf 54} (1994) 2756-2757.
 
 \bibitem{Solc71a}
 K. \u Solc,  Shape of a Random Flight Chain. {\it  J. Chem. Phys.} {\bf 55} (1971)  335-344. 

 \bibitem{Vincent96}
R.V.V. Vincent and N.A. Hill,  Bioconvection in a
suspension of phototactic algae. {\it  J. Fluid Mech.} { \bf 327} (1996) 343-371.

 
 \bibitem{Wang17} 
 Q. Wang et al.,  
Recent advances on smart TiO$_2$ nanotube platforms for sustainable drug delivery applications.
 {\it  Int. J. Nanomedicine } { \bf 12} (2017) 151-165. 
 
 \bibitem{Yurk17}
 B.P. Yurk,   Homogenization analysis of invasion dynamics in
heterogeneous landscapes with differential bias and
motility. {\it  J. Math. Biol.} { \bf 77} (2017) 27-54.

\bibitem{Weber13}
M. Weber et al., 
Interstitial dendritic cell guidance by haptotactic chemokine gradients. {\it  Science}
{\bf 339}
 (2013) 
328-332.

\bibitem{Wei97}
G. Wei and X. Zhu,  Shapes and sizes of arbitrary random walks at $O(1/d^3)$
II. Asphericity and prolateness parameters. {\it  Physica A} {\bf 237} (1997) 423-440.


\bibitem{Zakharchenko18}
A. Zakharchenko,  N. Guz, A.M. Laradji, E. Katz, and S.  Minko,   Magnetic field remotely controlled selective biocatalysis.
{\it   Nature Catalysis} 
{\bf 1} 
(2018) 
73-81. 



\end{thebibliography}
\end{document}